\documentclass[12pt]{article}
\def\today{15 July, 1996}
\newcommand{\cB}{{\cal B}}
\newcommand{\sde}{{\sc sde}}
\newcommand{\scm}{{\sc scm}}
\newcommand{\ord}[1]{{\cal O}\left(#1\right)}
\newcommand{\Z}[1]{Z^{(#1)}}
\newcommand{\V}[1]{V^{(#1)}}
\newcommand{\g}[1]{g^{(#1)}}
\newcommand{\eps}{\epsilon}

\usepackage{epsfig}
\epsfxsize=\textwidth
\renewcommand{\includegraphics}[1]{\epsfbox{#1}}

\begin{document}

\title{\bf On the low-dimensional modelling of
Stratonovich stochastic differential equations}
\author{Chao Xu \and A.J.~Roberts\thanks{Both at the Department of
Mathematics \& Computing, University of Southern Queensland, Toowoomba,
Queensland 4350, Australia. E-mail: {\tt chao@usq.edu.au} and {\tt
aroberts@usq.edu.au}}}
\maketitle

\begin{abstract}
We develop further ideas on how to construct low-dimensional models of
stochastic dynamical systems.  The aim is to derive a consistent and
accurate model from the originally high-dimensional system.  This is done
with the support of centre manifold theory and techniques.  Aspects of
several previous approaches are combined and extended: adiabatic
elimination has previously been used, but centre manifold techniques
simplify the
noise by removing memory effects, and with less algebraic effort
than normal forms; analysis of associated Fokker-Plank equations replace
nonlinearly generated noise processes by their long-term equivalent white
noise.  The ideas are developed by examining a simple dynamical system
which serves as a prototype of more interesting physical situations.
\end{abstract}


\section{Introduction}

Centre manifold theory is increasingly recognised as providing a rational
route to the low-dimensional modelling of high-dimensional dynamical
systems.  Applications of the techniques have ranged over, for example,
triple convection \cite{Arneodo85c}, feedback control \cite{Boe89},
economic theory \cite{Chiarella90}, shear dispersion \cite{Mercer90},
nonlinear oscillations \cite{Shaw93}, beam theory \cite{Roberts93}, flow
reactors \cite{Balakotaiah92}, and the dynamics of thin fluid films
\cite{Roberts94c}.  New insights given by the centre manifold picture
enable one to not only derive the dynamical models, but also to provide
accurate initial conditions \cite{Roberts89b,Cox93b}, boundary conditions
\cite{Roberts92c}, and, particularly relevant to this paper, the treatment
of forcing \cite{Cox91}.

The above developments have all taken place in the context of deterministic
dynamical systems. However, in practice, we want to apply the same concepts
in a noisy environment. Just one example of interest is the along-stream
dispersion of a contaminant in a turbulent river---the turbulence can only
reasonably be modelled by utilising stochastic factors. Thus it is vital
that we address the problem of constructing low-dimensional models of
stochastic differential equations ({\sde}s) using centre manifold ideas and
techniques.

Boxler \cite{Boxler89} has put the subject on a firm base by proving the
existence of centre manifolds for {\sde}s.  Further, these are proven
relevant as models for the full system; for example the reduced system of
evolution equations do indeed predict the stability of a degenerate fixed
point.  However, the analysis is very sophisticated and the results, as
seen in the given examples, involve infinite sums which seem impractical in
most applications.  In Sections~\ref{sdd} and~\ref{sred} we show how to
straightforwardly construct simpler versions of such centre manifolds.

Earlier work by Sch\"oner \& Haken \cite{Schoner86,Schoner87} used the
concepts of slaving and adiabatic elimination to develop one approach to
constructing low-dimensional models of {\sde}s.  Their analysis initially
appears
very similar, in Section~\ref{sdd} we derive exactly equivalent
results, but there are useful improvements through the centre manifold
viewpoint.  One immediate benefit, in direct contrast to more traditional
methods such as that of multiple scales, is that one can be much more
flexible about the ordering of various effects in the model.  Another is
that we present the analysis much more cleanly as an algorithm rather
than a hierarchy of formulae; one that can be easily adapted
\cite[\S5]{Cox91} to the vast majority of problems where it is inconvenient
to change basis in order to separate the ``master modes'' from the ``slave
modes.''

Further, the work of Cox \& Roberts \cite{Cox91,Roberts89b} showed how to
use flexibility inherent in the centre manifold approach to remove
memory effects of time-dependent forcing from the evolution on the centre
manifold (the forcing is
then purely local in time), unlike the approach of Sch\"oner \& Haken
\cite{Schoner86,Schoner87}.  However, only the leading order effect of the
forcing was considered.  In a separate line, and following the work of
Coullet {\em et al} \cite{Coullet85}, Sri Namachchivaya \& Lin
\cite{Srinamachchi91} used the techniques of normal form coordinate
transformations, to show how to extract a low-dimensional stochastic centre
manifold.  The normal form transformation has the same flexibility
recognised by Cox \& Roberts and consequently also demonstrates the
simplification in the evolution that may be attained.  Thus, in
Section~\ref{sred} we naturally extend the approach of Cox \& Roberts
\cite{Cox91} to significantly simplify the models of Sch\"oner \& Haken
\cite{Schoner86}. In large, physically interesting problems, dimensional
reduction is achieved via the approach described herein with much less
algebra than is needed by the normal form approach.

Lastly, preceding authors have presented their results in terms of noise
processes that have the same short-time scale of the transients
that we set out to eliminate by the modelling.  This is inconsistent.
Instead we propose that the irreducible noise processes in the model should
be further simplified by only considering them on the long-time scale of
the centre manifold evolution.  This is obtained, Section~\ref{sfp}, from a
centre manifold analysis of the Fokker-Planck equation for a noisy
dynamical system.  However, as also pointed out by Sri Namachchivaya \& Lin
\cite{Srinamachchi91}, Knobloch \& Wiesenfeld's \cite{Knobloch83} centre
manifold analysis of Fokker-Planck equations needs to be made more
systematic.  For the relatively simple {\sde}s arising here, the analysis
is straightforward and may be computed to high-order.  We deduce that these
noises may be replaced by appropriately chosen drift and white noise terms.

In summary, in this paper we bring the above separate threads together. We
show how to combine the best features of each to develop an efficient and
practical approach for low-dimensional modelling of dynamical systems.
The ideas are developed within the context of a simple modelling
problem.  We do this to pinpoint what is achieved and what needs to be
done, without obscuring the issues with complicating algebraic generalities.

Throughout, we adopt the Stratonovich interpretation of {\sde}s.

\section{Direct reduction}
\label{sdd}

In this section we demonstrate that the standard form of the algorithm to
find a forced centre manifold, as derived by Cox \& Roberts \cite{Cox91},
when extended to higher order in the forcing amplitude and applied to
Stratonovich {\sde}s produces the same results as obtained by Sch\"oner \&
Haken \cite{Schoner86} using the slaving principle (the similarity in the
basic approaches has previously been commented on, for example by Wunderlin
\& Haken \cite{Wunderlin81}).  However, as in the method of normal forms
\cite[\S4.2]{Srinamachchivaya90} and unlike simpler methods, the centre
manifold approach easily caters for nilpotent linearisations (for example,
the smallness assumption on the linearisation of the master modes of the
slaving principle, that $\Lambda_u\sim\delta$ in (3.2) of \cite{Schoner86},
is unnecessarily restrictive), and is readily generalised to construct
accurate invariant manifolds \cite{Roberts89}.  Further, the great
advantage the centre manifold approach has over the normal form procedure
of Sri Namachchcivaya \& Lin \cite[\S4.2]{Srinamachchivaya90} is that one
need only compute the low-dimensional centre manifold; unlike normal form
calculations, no algebra is done to describe the uninteresting details of
the large number, possibly infinite in number
\cite[e.g.]{Mercer90,Roberts94c}, of exponentially decaying modes.

The dynamical model we develop in this section provides basic results for
later comparison and improvement.

\subsection{Formal elimination near a pitchfork bifurcation}

Consider the following example {\sde}
discussed by Sch\"oner \& Haken \cite[\S5]{Schoner86}
\begin{eqnarray}
dx&=&(\alpha x-axy)dt+F_xdW_1\,,\label{eqx} \\
dy&=&(-\beta y+bx^2)dt+F_ydW_2\,, \label{eqy}
\end{eqnarray}
where $W_1$ and $W_2$ are independent Wiener processes, and $\alpha$,
$\beta$, $a$ and $b$ are real constants (with $a$ and $b$ having the same
sign).  In the absence of the stochastic forcing this dynamical system
undergoes a pitchfork bifurcation as the parameter $\alpha$ crosses 0; here
we may be interested in investigating effects upon the bifurcation of
the additive noise.

Near the bifurcation, for small $\alpha$, $x(t)$ evolves slowly whereas
$y(t)$ decays exponentially to zero.  Thus $x$ is the order parameter and
$y$ the slaved process.  More rigorously, we append the dynamical equation
$d\alpha=0dt$ and claim that Theorem~6.1 of Boxler \cite{Boxler89} asserts
that there exists a local stochastic centre manifold
for~(\ref{eqx}--\ref{eqy}) of the form $y=h(t,x,\alpha)$ for $x$ and
$\alpha$ small enough.

Thus we assume the stochastic centre manifold ({\scm}) is
\begin{equation}
	y=h(t,x)\,,
	\label{yhass}
\end{equation}
where the dependence upon $\alpha$ is implicit, and where we take the
slaved variable to explicitly depend on only two variables: time; and the
order parameter $x$.  Sch\"oner \& Haken \cite[Eq.(2.1)]{Schoner86} instead
used many variables, $y=h\left(t,x,Z_\nu\right)$, where $Z_\nu(t)$ denotes
a large number of as yet unknown noise processes which appear in the
low-dimensional model.  Here we considerably simplify the algebra by
including, albeit implicitly, the effects of such noise through the time
variable $t$.  Consequently, integrals with respect to time are performed
treating $x$ as a constant, but integrating any noise.

Differentiate~(\ref{yhass}) with respect to $t$ to
\begin{eqnarray}
dy=\frac{\partial h}{\partial t}dt+\frac{\partial h}{\partial
x}dx\,.\label{eqdx}
\end{eqnarray}
Substituting (\ref{eqdx}), (\ref{eqy}) is written in the following form:
\begin{eqnarray}
\left[\frac{\partial }{\partial t}dt+\beta I
dt\right]h&=&bx^2dt+F_ydW_2-\frac{\partial h}{\partial x}dx\,. \label{dyyy}
\end{eqnarray}
Substituting for $dx$ from~(\ref{eqx}) we obtain:
\begin{eqnarray}
{\cB}h
&=&bx^2dt+F_ydW_2-\frac{\partial h}{\partial x}\left[(\alpha
x-axh)dt+F_xdW_1\right]\,,  \label{eqLL}
\end{eqnarray}
where
\[ {\cB}=\frac{\partial }{\partial t}dt+\beta I\,dt\,.
\]
This equation is to be solved for $h$, the {\scm}, albeit approximately.

In the following we solve a sequence of problems of the form
\begin{eqnarray}
{\cB}g=f_1dt+f_2dW\,,\label{lg}
\end{eqnarray}
where $f_1(t,x)$ and $f_2(t,x)$ are known functions, and we need to
determine the process $g(t,x)$.
We may deduce\cite[Eq.(2.20)]{Schoner86}
\[ g=e^{-\beta t}\star(f_1dt+f_2dW)
=\int_{-\infty}^t e^{-\beta(t-\tau)}\left(f_1d\tau+f_2dW(\tau)\right)\,,
\]
where ``$\star$'' denotes the given convolution. Note these convolution
integrals are done keeping $x$ constant as that is implicit in the notation
$\partial/\partial t$ of time differentiation, whereas all noise processes
appearing in the integrand must be integrated.

\subsection{Asymptotic solution}

To simplify the solution procedure we define $\delta$ to be
\[
\delta^2=\|x(t)\|^2+|F_x|+|F_y|+|\alpha|\,,
\]
so that $\delta$ is a quantitative measure of the overall size of the solution
and contributing terms. In effect, for small $\delta$ the following
scaling relations hold:
\begin{eqnarray}
&&x=\ord{\delta}\,,\qquad
\alpha,\, F_x,\, F_y\ \mbox{are}\ \ord{\delta^2}\,,\nonumber
\end{eqnarray}
whereas the other constants are of order~0 in $\delta$.  In principle, the
centre manifold formalism allows us to expand the solution in terms of the
parameters independently.  However, as discussed in \cite{Roberts88a}, by
using one ordering parameter $\delta$ we obtain the same set of terms, it
is just that they appear in the asymptotic expansion in a regimented order.
Thus we expand
\begin{eqnarray}
h&\sim&\sum_{m=2}^{\infty}h^{(m)}(t,x)\,,  \label{yt}
\end{eqnarray}
where $h^{(m)}$ contains all terms of order $m$ in $\delta$.
Substituting into (\ref{eqLL}), we obtain
\begin{eqnarray}
{\cB}\sum_{m=2}^{\infty}h^{(m)}(t,x)
&\sim&bx^2dt+F_ydW_2\nonumber\\
&&-\sum_{m=2}^{\infty}{\frac{\partial h}{\partial
x}}^{(m)}\left[\left(\alpha
x-ax\sum_{m=2}^{\infty}h^{(m)}\right)dt+F_xdW_1\right]\,. \label{itc2}
\end{eqnarray}
\begin{itemize}
\item Extracting all terms of order $m=2$,
\begin{eqnarray}
h^{(2)}(t,x)&=&\cB^{-1}\left[bx^2dt+F_ydW_2\right]\nonumber\\
&=&e^{-\beta t}\star \left[bx^2dt+F_ydW_2\right]\label{m21}\\
&=&\frac{b}{\beta}x^2+F_y\Z2\,,\nonumber
\end{eqnarray}
where
\begin{equation}
\Z2(t)=\int_{-\infty}^te^{-\beta(t-\tau)}dW_2(\tau)
=e^{-\beta t}\star dW_2\,.
	\label{aZ2}
\end{equation}

\item Terms of order $m=3$ similarly lead to
\begin{eqnarray}
h^{(3)}(t,x)\label{m31}
&=&-\frac{2b}{\beta}xF_x\Z3\,,
\end{eqnarray}
where
\begin{equation}
	\Z3=e^{-\beta t}\star dW_1\,.
	\label{aZ3}
\end{equation}

\item Terms of order $m=4$ give
\begin{eqnarray}
h^{(4)}(t,x)&=&-\frac{2\alpha
bx^2}{\beta^2}+\frac{2ab^2x^4}{\beta^3}+\frac{2b}{\beta}{F_x}^2\Z4_1
+\frac{2abx^2}{\beta}F_y\Z4_2\,,\label{m44}
\end{eqnarray}
where
\begin{eqnarray}
\Z4_1=e^{-\beta t}\star \Z3dW_1\,,
&\quad\mbox{and}\quad &
\Z4_2=e^{-\beta t}\star \Z2dt\,.\label{aZ4}
\end{eqnarray}

\item Terms of order $m=5$ give
\begin{eqnarray}
h^{(5)}(t,x)&=&\frac{2\alpha
b}{\beta}xF_xZ_3^{(5)}-\frac{6ab^2}{\beta^2}x^3F_xZ_3^{(5)}
-\frac{2abx}{\beta}F_xF_yZ_4^{(5)}\label{m54}\\
&&+\left(\frac{4\alpha
bx}{\beta^2}-\frac{8ab^2x^3}{\beta^3}\right)F_xZ_1^{(5)}
-\frac{4abx}{\beta}F_xF_yZ_2^{(5)}\,,\nonumber
\end{eqnarray}
where
\begin{eqnarray}
Z_1^{(5)}=e^{-\beta t}\star dW_1\ \left(=\Z3\right)\,,&\quad&
Z_2^{(5)}=e^{-\beta t}\star \Z4_2dW_1\,,\label{aZ5}\\
Z_3^{(5)}=e^{-\beta t}\star \Z3dt\,,&\quad&
Z_4^{(5)}=e^{-\beta t}\star \Z2\Z3dt\,.\nonumber
\end{eqnarray}
\end{itemize}
Clearly this procedure may be continued indefinitely to derive, in
principle, arbitrarily high-order asymptotic approximations to the
stochastically forced centre manifold.

After substituting the approximate $y=h(t,x)$ into the $x$
equation, we get the following stochastic evolution equation, accurate to
6th order in $\delta$:
\begin{eqnarray}
dx&\sim&F_xdW_1+\left\{
\left[\alpha x-\frac{ab}{\beta}x^3-aF_y\Z2x\right]
+\left[\frac{2ab}{\beta}F_x\Z3x^2\right]
\right.\nonumber\\&&
+\left[\frac{2a\alpha b}{\beta^2}x^3
-\frac{2a^2b^2}{\beta^3}x^5-\frac{2a^2b}{\beta}F_y\Z4_2x^3
-\frac{2ab}{\beta}{F_x}^2\Z4_1x\right]
\nonumber\\&&
+\left[-\frac{2a\alpha b}{\beta}F_xZ_3^{(5)}x^2
+\frac{6a^2b^2}{\beta^2}F_xZ_3^{(5)}x^4
+\frac{2a^2b}{\beta}F_xF_yZ_4^{(5)}x^2
\right.\label{m61}\\
&&\left.\left.
\quad-\left(\frac{4a\alpha
b}{\beta^2}x^2-\frac{8a^2b^2}{\beta^3}x^4\right)F_xZ_1^{(5)}
+\frac{4a^2b}{\beta}F_xF_yZ_2^{(5)}x^2\right]\right\}dt\,.
\nonumber
\end{eqnarray}
This last {\sde} describes the low-dimensional
evolution relevant to the noisy pitchfork bifurcation. As mentioned
earlier, these results are identical to those of Sch\"oner \& Haken
\cite[Eq.(5.15)]{Schoner86}, but the derivation is shorn of unnecessary
complicating detail and in the framework of the more powerful centre
manifold approach.

\section{Simplify the noise processes}
\label{sred}

One of the remarkable features of the preceding analysis is the appearance
of many noise processes in the low-dimensional model~(\ref{m61}).  In this
example there are 8 new noises up to fifth order; in interesting physical
examples there would be many more corresponding to each of the neglected
transient modes (see (28--29) in \cite{Coullet85} for example).  Further,
each of these noise processes involves a memory, they are each a
convolution integral over the past history of a noise source.  In the
context of simply forced dynamical systems, Cox \& Roberts \cite{Cox91}
showed how to adapt the centre manifold procedure to eliminate the memory
effect and consequently simplify the forcing in low-dimensional models;
however, the analysis was only linear in the amplitude of the forcing.  In
a similar spirit, Coullet {\em et al\/} \cite{Coullet85} and later Sri
Namachchivaya \& Lin \cite{Srinamachchi91} examined additive and
linearly-multiplicative noise, using the normal form transformation to
limit the complexity of not only the deterministic terms, but also to
simplify the noise on the centre and the stable manifolds.

In this section we show that the approach of Cox \& Roberts \cite{Cox91},
in conjunction with new ideas on the removal of resonant stochastic terms,
leads directly to a simple model on the centre manifold; thus simplifying
the methods and results compared with those obtainable with the methods of
previous authors.  The two critical steps are to allow some flexibility in the
parameterization of the centre manifold and to examine the stochastic
resonances directly rather than in Fourier space.  In doing so, as
discussed in \cite{Cox91}, we implicitly cater for the nature of the
evolution near to the centre manifold and consequently discover how to
treat the stochastic forcing in a simple manner.  Again we introduce the
ideas as applied to the simple system~(\ref{eqx}--\ref{eqy}).

\subsection{Reparameterise the centre manifold}

Assume the centre manifold is described by
\begin{eqnarray}
x&=&s+\xi(t,s)\,,                         \label{tcent1}\\
y&=&\eta(t,s)\,,                          \label{tcent}\\
ds&=&g(t,s)  =g_0(t,s)dt+g_1(t,s)dW_1+g_2(t,s)dW_2\,,  \label{tcent2}
\end{eqnarray}
where $s$ parameterizes the centre manifold and hence $s(t)$ describes the
evolution of the low-dimensional model.  Note that (\ref{tcent1}) is a near
identity
transformation of $x$ and we use the flexibility in $\xi$ to eliminate as
far as possible unnecessary noise.  The analysis is in the spirit of normal
forms; however, here we directly construct a normal form on the centre manifold
without going through the laborious task of finding a normal form of the
entire system.  Substitute (\ref{tcent}) and (\ref{tcent1}) into the $y$
and $x$ evolution equations, (\ref{eqy}) and (\ref{eqx}), to get:
\begin{eqnarray}
dy&=&\frac{\partial \eta}{\partial t}dt+\frac{\partial \eta}{\partial
s}ds=\left[-\beta \eta+b(s+\xi)^2\right]dt+F_ydW_2\,,\label{tcee1}\\
dx&=&\frac{\partial \xi}{\partial t}dt
+\left(1+\frac{\partial \xi}{\partial s}\right)ds=\left[\alpha s+\alpha
\xi-a(s+\xi)\eta\right]dt+F_xdW_1\,,\label{tcee2}
\end{eqnarray}
and hence, using~(\ref{tcent2}),
\begin{eqnarray}
\cB\eta
&=&b(s^2+2s\xi+\xi^2)dt+F_ydW_2-\frac{\partial \eta}{\partial
s}g\,,\label{tcee3}\\
g&=&-\frac{\partial \xi}{\partial t}dt-\frac{\partial \xi}{\partial s}g
+\left[\alpha s+\alpha \xi-as\eta-a\xi\eta\right]dt+F_xdW_1\,.\label{tcee4}
\end{eqnarray}
Substitute the asymptotic expansions
\begin{eqnarray*}
	\eta & \sim & \eta^{(2)}+\eta^{(3)}+\eta^{(4)}+\cdots\,,  \\
	\xi & \sim & \xi^{(2)}+\xi^{(3)}+\xi^{(4)}+\cdots\,,  \\
	g & \sim & g^{(2)}+g^{(3)}+g^{(4)}+\cdots\,,
\end{eqnarray*}
where, as before, the superscript ${\ }^{(n)}$ denotes quantities of
order~$n$ in $\delta$ (note that $s=\ord{\delta}$).
After comparison of the order of the terms appearing in (\ref{tcee3}) and
(\ref{tcee4}), we obtain the following equations and their solution.

\subsection{Order 2: simple resonance}

Terms of order $m=2$ give:
\begin{eqnarray}
\cB \eta^{(2)}&=&bs^2dt+F_ydW_2\,,
\quad\mbox{and}\quad
g^{(2)}=-{\frac{\partial\xi}{\partial t}}^{(2)}dt+F_xdW_1\,.\label{tcee5}
\end{eqnarray}
It is not reasonable to use the freedom in $\xi^{(2)}$ to eliminate
$dW_1$ in $g^{(2)}$ in order to simplify the model {\sde} $ds=g$, therefore
\begin{eqnarray}
\xi^{(2)}=0\,,
\quad&&
\eta^{(2)}=\frac{b}{\beta}s^2+F_y\Z2\,,\label{tcee6}
\quad\mbox{and}\quad
g^{(2)}=F_xdW_1\,,
\end{eqnarray}
where $\Z2=e^{-\beta t}\star dW_2$ as in~(\ref{aZ2}).  If one attempts to
use $\xi^{(2)}$ to eliminate the noise $F_xdW_1$, then one is lead to
$\xi^{(2)}=W_1$ which over long-times grows like $t^{1/2}$.  Such secular
growth, albeit stochastic, leads to an unallowable non-uniform convergence
in time: eventually $\xi^{(2)}$ would no longer be a small perturbation to
the leading-order term in $x=s+\xi(t,s)$.  Thus the component in $dW_1$
must remain in $g$.  This sort of direct argument about long-time
behaviour, rather than resonance in Fourier space, determines which terms
we may eliminate through the reparameterization (normal form
transformation) of the centre manifold.

\subsection{Order 3: coloured noise}

Terms of order $m=3$ lead to:
\begin{eqnarray*}
\cB \eta^{(3)}
=-\frac{2b}{\beta}sF_xdW_1\,,
&\quad\mbox{and}\quad&
g^{(3)}=\left(\alpha s-\frac{ab}{\beta}s^3
-{\frac{\partial \xi}{\partial t}}^{(3)}-asF_y\Z2\right)dt\,.
\end{eqnarray*}
$\xi^{(3)}$ cannot be used to eliminate either the $\alpha s$ nor the $s^3$
terms as they have non-zero mean (over $t$, keeping $s$ constant), and
hence would generate secular growth if they were included in $\xi^{(3)}$.
Also, we would like ${\frac{\partial \xi}{\partial t}}^{(3)}+asF_y\Z2=0$,
namely $\xi^{(3)}=-asF_y\int \Z2dt$, in order to
simplify the above equation for $g^{(3)}$; however, this is not possible as
we now explain.  From~(\ref{aZ2}), the coloured noise $\Z2$ satisfies
the {\sde}
\begin{equation}
	d\Z2=-\beta \Z2dt+dW_2\,.
	\label{dZ2}
\end{equation}
Thus over long-times, on the time-scales we are interested in for the
low-dimensional model, namely those that are much longer than $1/\beta$,
$\Z2=\ord{t^{-1/2}}$ as $dt=\ord{t}$ and $dW_2=\ord{t^{1/2}}$.
Consequently, $\int \Z2dt=\ord{t^{1/2}}$ is unallowably secular, and
hence it would appear, as in \cite{Coullet85,Srinamachchi91}, that the
$\Z2dt$ noise must remain in the model {\sde} $ds=g$.

Nonetheless, as in the analysis of deterministic forced systems by Cox \&
Roberts \cite{Cox91}, we can usefully simplify the model {\sde} by removing
the ``colour'' of the noise, that is we avoid the memory intrinsic to $\Z2$ and
processes like it.  But first a digression.

One of the benefits of only considering the dynamics on the centre
manifold, apart from its low-dimensionality, is that often the evolution
is intrinsically slow; and even when the evolution is characterised by fast
oscillation, as in a Hopf bifurcation \cite{Coullet85} for example, a
normal form transformation renders the evolution in terms of slowly-varying
amplitudes.  Thus we expect to only focus our attention on the long-time
scales of the model---for example, one may take long time-steps in a
numerical solution---and we do not need to resolve any details of the
rapidly decaying transients.  But the form of the
evolution~(\ref{m61}) on the stochastic centre manifold involves noise
processes $Z^{(\nu)}(t)$ which are coloured on the fast-time scale of the
decaying transients.  This can be seen clearly in~(\ref{dZ2}), and
similarly for the other noises introduced in the previous section.
Here each noise process is dominated by the decay on the fast-time scale of
$1/\beta$. Precisely the same effect can be seen in the normal form results
of Sri Namachchivaya \cite{Srinamachchi91}, Eqs.(14--15) in general and
Eq.(39) in particular, where the noise on the centre manifold modes,
$U_{cs}$ in general, decays according to the linearised operator of the
decaying modes. It is also seen in Eq.(29) of Coullet {\em et al\/}
\cite{Coullet85}.  However, the latter also point out that on the
long-time scales of the model, a coloured noise may be approximated by a
white noise, here $\Z2dt\approx dW_2/\beta$: a result easily identified
from a balance of terms in~(\ref{dZ2}).  We now proceed to show how to
turn this simplifying approximation into an exact result.

Rewrite the {\sde}~(\ref{dZ2}) for $\Z2$ in the form
\[ \Z2dt=-\frac{d\Z2}{\beta}+\frac{dW_2}{\beta}\,,
\]
whence it follows that we can replace $\Z2dt$ by the above in the
right-hand side of the equation for $g^{(3)}$ to obtain
\[ g^{(3)}=\left(\alpha s-\frac{ab}{\beta}s^3\right)dt
-{\frac{\partial \xi}{\partial t}}^{(3)}dt+\frac{a}{\beta}sF_yd\Z2
-\frac{a}{\beta}sF_y dW_2\,.
\]
Now the term in $d\Z2$ can be absorbed into $\xi^{(3)}$, leaving just the
white noise $dW_2$ in $g^{(3)}$ for the model {\sde}.
Thus
\begin{eqnarray}
	 & \xi^{(3)}=\frac{a}{\beta}sF_y\Z2\,,\quad
\eta^{(3)}=-\frac{2b}{\beta}sF_x\Z3\,,
 &
	\nonumber \\
	\mbox{and} & g^{(3)}=\left(\alpha s-\frac{ab}{\beta}s^3\right)dt
-\frac{a}{\beta}sF_y dW_2\,,
 &
	\label{tcee7}
\end{eqnarray}
where $\Z3=e^{-\beta t}\star dW_1$ as before in~(\ref{aZ3}).

In
summary, to third order we deduce the centre manifold is
\begin{eqnarray*}
	x & \sim & s+\frac{a}{\beta}sF_y\Z2\,,  \\
	y & \sim & \frac{b}{\beta}s^2+F_y\Z2
	-\frac{2b}{\beta}sF_x\Z3\,,  \\
	\mbox{s.t.}\quad
	ds & \sim & \left(\alpha s-\frac{ab}{\beta}s^3\right)dt
	+F_xdW_1 -\frac{a}{\beta}sF_y dW_2\,.
\end{eqnarray*}
This is an exact decomposition: to this order we have put all the
``colour'' into the description of the location and shape of the
stochastically forced centre manifold, leaving just pure white noise in the
model evolution equation.

\subsection{Order 4: nonlinearly generated noise}

Equating terms of order $m=4$ in~(\ref{tcee3}--\ref{tcee4}) give:
\begin{eqnarray*}
\cB \eta^{(4)}
&=&\frac{2ab}{\beta}s^2F_y\Z2 dt-\frac{2b}{\beta}s\left(\alpha
s-\frac{ab}{\beta}s^3\right)dt+\frac{2b}{\beta}F_x^2\Z3dW_1
\\&&\quad
+\frac{2ab}{\beta^2}s^2F_ydW_2\,,
\\
g^{(4)}&=&
-{\frac{\partial \xi}{\partial t}}^{(4)}dt
-\frac{a}{\beta}F_xF_y\Z2dW_1+\frac{2ab}{\beta^2}s^2F_xdW_1\,.
\end{eqnarray*}
The terms to the right of $\xi^{(4)}$ are resonant, $dW_1=\ord{t^{1/2}}$
and $\Z2dW_1=\ord{t^0}$, so that at least some component of these
noises has to be assigned to $g^{(4)}$.  At this stage we cannot think of
a useful transformation of the $\Z2dW_1$ contribution, and so must
choose
\begin{eqnarray}
\xi^{(4)}&=&0\,,\quad\mbox{and thus}
\nonumber\\
\eta^{(4)}&=&
-\frac{2b}{\beta^2}s\left(\alpha s-\frac{ab}{\beta}s^3\right)
+\frac{2ab}{\beta}F_ys^2\left(Z_2^{(4)}+\frac{\Z2}{\beta}\right)
+\frac{2b}{\beta}F_x^2\Z4_1
\,,\label{tcee10}\\
g^{(4)}&=&+\frac{2ab}{\beta^2}s^2F_xdW_1
-\frac{a}{\beta}F_xF_y\Z2dW_1\,,
\nonumber
\end{eqnarray}
where $\Z4_1=e^{-\beta t}\star \Z3dW_1$ and $\Z4_2=e^{-\beta
t}\star \Z2dt$ as before, (\ref{aZ4}).

A prime target for further research is to simplify the nonlinear
combination of $\Z2dW_1$; the difficulty is that this is not a normal
coloured noise term.  Its presence here shows that the typical assumption
of Gaussian noise is generally not justified in modelling nonlinear
dynamics.  However, in the next section we outline a
Fokker-Planck approach to the rational simplification of such nonlinear
noise processes.

\subsection{Order 5: noise induced drift}

Terms of order $m=5$ lead to:
\begin{eqnarray*}
\cB \eta^{(5)}&=&
\frac{2b}{\beta}(\alpha s-\frac{3ab}{\beta}s^3)F_x\Z3dt
-\frac{2ab}{\beta^2}sF_xF_y\Z3dW_2
\\
&&+\left[\frac{4b}{\beta^2}s(\alpha -\frac{3ab}{\beta}s^2)
-\frac{4ab}{\beta}sF_y\Z4_2
-\frac{2ab}{\beta^2}sF_y\Z2\right]F_xdW_1\,,
\\
g^{(5)}&=&-{\frac{\partial \xi}{\partial t}}^{(5)}dt
+\frac{2\alpha ab}{\beta^2}s^3dt
-\frac{2a^2b^2}{\beta^3}s^5dt
-\frac{2a^2b}{\beta}s^3F_y\left(\Z4_2+\frac{1}{\beta}\Z2\right)dt
\\
&&-\frac{2ab}{\beta}sF_x^2\Z4_1dt
-\frac{a^2}{\beta}sF_y^2{\Z2}^2dt
+\frac{4ab}{\beta^2}sF_x^2\Z3dW_1
+\frac{a^2}{\beta^2}sF_y^2\Z2dW_2
\,.
\end{eqnarray*}
As before we use $\xi^{(5)}$ to absorb as much of the evolution terms in
$g^{(5)}$.   Recall that
$\Z2dt=-\frac{1}{\beta}d\Z2+\frac{1}{\beta}dW_2$, so that
\[ {\Z2}^2dt=\Z2\left(\Z2dt\right)
=-\frac{1}{2\beta}d\left({\Z2}^2\right)+\frac{1}{\beta}\Z2dW_2\,,
\]
whereas from~(\ref{aZ4})
\begin{eqnarray*}
	\Z4_2dt & = & -\frac{1}{\beta}d\Z4_2+\frac{1}{\beta}\Z2dt
=-\frac{1}{\beta}d\Z4_2-\frac{1}{\beta^2}d\Z2+\frac{1}{\beta^2}dW_2\,,  \\
	\Z4_1dt & = & -\frac{1}{\beta}d\Z4_1+\frac{1}{\beta}\Z3dW_1\,.
\end{eqnarray*}
Thus we set
\begin{eqnarray}
\xi^{(5)}&=&
+\frac{2a^2b}{\beta^2}s^3F_y \left(\Z4_2+\frac{2}{\beta}\Z2\right)
+\frac{2ab}{\beta^2}sF_x^2 \Z4_1
+\frac{a^2}{2\beta^2}sF_y^2{\Z2}^2\,,\nonumber
\\
\eta^{(5)}&=&
\frac{2b}{\beta}\left(\alpha s-\frac{3ab}{\beta}s^3\right)
              F_x\left(\Z5_3+\frac{2}{\beta}\Z3\right)
-\frac{4ab}{\beta}sF_xF_y\Z5_2
-\frac{2ab}{\beta^2}sF_xF_y\Z5_5
\,,
\nonumber\\
g^{(5)}&=&\left(\frac{2\alpha ab}{\beta^2}s^3
-\frac{2a^2b^2}{\beta^3}s^5\right)dt
-\frac{4a^2b}{\beta^3}s^3F_ydW_2
+\frac{2ab}{\beta^2}sF_x^2\Z3dW_1\,,\label{tcee14}
\end{eqnarray}
where the new noise
\begin{equation}
	d\Z5_5=e^{-\beta t}\star\left(\Z2dW_1+\Z3dW_2\right)\,.
	\label{nZ5}
\end{equation}
A factor of interest in $g^{(5)}$ is the non-Gaussian noise $\Z3dW_1$
which has non-zero mean and predicts a long-term ``drift'' in the
low-dimensional model of the order of the square of the stochastic
forcing.  This term, being linearly multiplicative in $s$ will contribute to a
destabilisation of the origin.

\subsection{Asymptotic model}

Therefore, we obtain $ds$ to fifth order as follows:
\begin{eqnarray}
ds&\sim&\left(\alpha s-\frac{ab}{\beta}s^3+\frac{2\alpha
ab}{\beta^2}s^3-\frac{2a^2b^2}{\beta^3}s^5\right)dt
\nonumber\\&&
+\left(1+\frac{2ab}{\beta^2}s^2\right)F_x dW_1
-\left(\frac{a}{\beta}s+\frac{4a^2b}{\beta^3}s^3\right)F_y dW_2
\label{tcee15}\\&&\nonumber
-\frac{a}{\beta}F_xF_y\Z2dW_1
+\frac{2ab}{\beta^2}sF_x^2\Z3dW_1
\,.
\end{eqnarray}
This has significantly simpler structure than the fifth order version of
Sch\"oner \& Haken that we derived in the previous section,
namely~(\ref{m61}) truncated to
\begin{eqnarray}
dx&\sim&F_xdW_1+\left\{
\left[\alpha x-\frac{ab}{\beta}x^3-aF_y\Z2x\right]
+\left[\frac{2ab}{\beta}F_x\Z3x^2\right]
\right.\nonumber\\&&\left.
+\left[\frac{2a\alpha b}{\beta^2}x^3
-\frac{2a^2b^2}{\beta^3}x^5-\frac{2a^2b}{\beta}F_y\Z4_2x^3
-\frac{2ab}{\beta}{F_x}^2\Z4_1x\right]
\right\}dt\,.
\nonumber
\end{eqnarray}
There are only two new noise processes in the model, instead of four, and
the two noises which are new are obtained from lower order {\sde}s.
Instead we see in~(\ref{tcee15}) a richer structure directly in terms of
the noise from the original system.

\section{Fokker-Planck analysis of nonlinear noise}
\label{sfp}

In this section we turn to the appearance of the noise processes $\Z3dW_1$
and $\Z2dW_1$ in the low-dimensional model~(\ref{tcee15}).  The computation
of $\Z2$ and $\Z3$, via the convolutions in (\ref{aZ2}) and (\ref{aZ3}),
necessarily takes place on the fast time-scale $1/\beta$, and must
therefore be a major hindrance to the use of the model over long times.
Such fast time-scales in the noise are not apparent in the centre manifold
analysis of Fokker-Planck equations by Knobloch \& Wiesenfeld
\cite{Knobloch83}.  In a pitchfork bifurcation, for example, the evolution
on the centre manifold of a Fokker-Planck equation
\cite[Eq.(3.22)]{Knobloch83} is equivalent to the stochastic differential
equation
\[ du=\left(\mu u+au^3\right)dt+\sigma d\xi\,,
\]
where $d\xi(t)$ represents Gaussian white noise.  In simpler systems one
can carry out an analysis of the Fokker-Planck equations to high order and
see that while the effective noise may be non-Gaussian, there is no remnant of
the fast time-scale.  We show this here, and
then suggest how to model $\Z3dW_1$ and $\Z2dW_1$.

We propose to replace these nonlinearly generated noises with new
independent Wiener processes with the same long-term statistics.  In
essence we construct a {\em weak\/} low-dimensional model by these
arguments---fidelity with the original dynamics can only be assured via
the undesired evaluation of fast-time integrals.

\subsection{Self excited drift}

In the {\sde}~(\ref{tcee15}), we would like to replace the factor
$\Z3dW_1$ by a $dw$ where $dw$ represents some simple noise process
which could be sampled over relatively large times.  That is, using the
definition of $\Z3$, we seek to understand something of the evolution of
$w(t)$ in
\[
	dw = \Z3dW_1\,,  \quad
	d\Z3 = -\beta\Z3dt+dW_1\,.
\]
This is in the form of a {\sde} with one neutral mode $w$ and one
exponentially decaying mode $\Z3$.  Thus we may apply centre manifold
techniques to derive a model {\sde} for the mode $w(t)$.  However, if this
is done using the techniques described earlier, then one finds exactly the
same type of noise process in the results.  The above pair of coupled
{\sde}s is in some sense canonical in that it is irreducible using the
previous techniques.  Instead we turn to the Fokker-Planck equation; we
must do something different because the results, equation~(\ref{fpsde}),
involve the surd $\sqrt\beta$ which is impossible to obtain via the direct
application of the algebra of centre manifold techniques.

For convenience we rewrite the {\sde} as
\begin{equation}
	dx = \eps ydW_1\,,
	\quad
	dy = -\beta ydt+\eps dW_1\,,
	\label{fpa}
\end{equation}
where, in this subsection {\em only}, $x=w$ and $y=\Z3$.  The ``small''
parameter $\eps$ has been introduced to order the terms in the asymptotic
analysis; eventually we set $\eps=1$ to recover approximate results for the
original process.  The Fokker-Planck equation of this Stratonovich {\sde}
is
\begin{equation}
	\partial_t u=
	\frac{\eps^2}{2}\partial_x u
	+\beta\partial_y\left(yu\right)
	+\frac{\eps^2}{2}\left\{y^2\partial_{xx}u +2\partial_{xy}(yu)
	   +\partial_{yy}u \right\}\,,
	\label{fpb}
\end{equation}
for the probability density function $u(x,y,t)$.  We seek a model for the
long-term evolution of $x(t)$ via the derivation of a Fokker-Planck
equation for the probability density function
\begin{equation}
	p(x,t)=\int_{-\infty}^\infty u(x,y,t)\,dy\,.
	\label{fpc}
\end{equation}
Heuristically, the term $\beta\partial_y\left(yu\right)$ in~(\ref{fpb})
concentrates $u$ upon $y=0$, but this is countered by the noise induced
diffusion term $\frac{\eps^2}{2}\partial_{yy}u$.  Thus to leading order,
$u$ is Gaussian in $y$---but very localised if $\eps$ is small.  Then the
other terms in the Fokker-Planck equation cause a long-term drift and
spread in the $x$ direction which we describe in terms of $p(x,t)$.

We use centre manifold techniques to find the long-term evolution of $p$.
Because of the symmetry and simplicity of~(\ref{fpa}) we can analyse the
Fokker-Planck equation somewhat more straightforwardly than Knobloch \&
Wiesenfeld
\cite{Knobloch83}.  First, scale
\[ y=\eps Y\,,\quad u=\frac{1}{\eps}U(x,Y,t)\,,
\]
so that with $Y$ we can resolve the Gaussian structure, and any
modifications, in $y$.  Then seek
\begin{eqnarray}
	U & = & V(Y,p) \sim\sum_{n=0}^\infty \eps^n\V{n}(Y,p)\,,
	\label{fpcm} \\
	\mbox{such that}\quad
	\partial_t p & = & g(p)	\sim\sum_{n=0}^\infty \eps^n\g{n}(p)\,.
	\label{fpevol}
\end{eqnarray}
Substituting these into~(\ref{fpb}) and~(\ref{fpc}) gives a hierarchy of
equations to solve for $\V n$ and $\g n$.  Using computer algebra it is
straight-forward to find
\begin{eqnarray}
	U & \sim & G(Y)p
	\nonumber \\
	 &  & +\eps^2\left(-Y^2 +\frac{1}{2\beta}\right)G(Y)p_x
	\label{fpcm3} \\
	 &  & +\eps^4\left(\frac{1}{2}Y^4 -\frac{1}{4\beta}Y^2
	 -\frac{1}{4\beta^2}\right)G(Y)p_{xx}
	\nonumber \\
	 &  & +\eps^6\left(-\frac{1}{6}Y^6 +\frac{1}{8\beta^2}Y^2
	 +\frac{1}{4\beta^3}\right)G(Y)p_{xxx} \,,
	 \nonumber \\
	 \mbox{such that}\quad
	 p_t&\sim& -\frac{\eps^2}{2}p_x +\frac{\eps^4}{4\beta}p_{xx}
	 -\frac{\eps^6}{4\beta^2}p_{xxx}\,,
	\label{fpev3}
\end{eqnarray}
where $G(Y)=\sqrt{\beta/\pi}\exp\left(-\beta Y^2\right)$
is the leading-order Gaussian structure.

This last evolution equation~(\ref{fpev3}), when truncated to neglect terms
$o\left(\eps^4\right)$, is the Fokker-Planck equation of a {\sde} such as
\begin{equation}
	dx\sim\frac{\eps^2}{2}dt+\frac{\eps^2}{2\sqrt{\beta}}dW\,,
	\label{fpsde}
\end{equation}
where the noise process $dW$ may be considered essentially independent of
the original noise $dW_1$ because, on the long-time scales that we wish to
use these noises, the nonlinear process generating $x(t)$ summarises
quite different information about the details of $dW_1$ than that of the simple
cumulative sum seen in $W_1(t)$.  This is confirmed by numerical
simulations.
\begin{figure}
\centering\includegraphics{zdwa1.ps}
\caption{scatter plots of $\Delta (x-t/2)$ versus $\Delta W_1$ on two
time-scales:
left-hand plot for $\Delta t=2/\beta$; right-hand plot for $\Delta
t=16/\beta$.  The data is generated by public domain software, Gnans,
numerically integrating~(\protect\ref{fpa}) over $0\leq t\leq 10,000$ with
a time-step of $0.01$.}
\protect\label{fzdw1}
\end{figure}
Figure~\ref{fzdw1} shows that, on time scales long compared with the decay
time $1/\beta$, there is virtually no linear correlation between the noises
$dW_1$ and $dx-dt/2$; there are nonlinear correlations for smaller times,
e.g.~$\Delta t=2/\beta$, but these slowly disappear, as seen for $\Delta
t=16/\beta$.  Thus we propose to replace, in the low-dimensional
model~(\ref{tcee15}), the fast time-scale factors $\Z3dW_1$ by the above
$dx$ for $\eps=1$, namely by $\frac{1}{2}dt+\frac{1}{2}dW_3/\sqrt{\beta}$
where $W_3$ is a new independent Wiener process.  In essence, this
replacement summarises the drift and overall fluctuations in $\Z3dW_1$
while ignoring the other details of the nonlinear noise.

There are two issues to comment on.  Firstly, it is reasonable to
truncate~(\ref{fpev3}) to the first two terms because this is the lowest
order structurally stable approximation to the evolution of $p(x,t)$.
Further, although there are higher-order modifications, it is apparent that
$\g{2n}=\ord{\beta^{1-n}}$ so that, provided the decay of the ``slaved''
modes are rapid enough, these higher-order terms may be
neglected while maintaining accuracy.

Secondly, strictly speaking the asymptotic expansion~(\ref{fpcm3}) is not
uniformly valid and so could be in error; it is evident that $\V n\propto
Y^nG(Y)$ and so $\V n\gg \V{n-1}$ for large enough $|Y|$.  However, instead
of solving for all $Y$ we may restrict analysis to
$|Y|<\Psi=\ord{\eps^{-1/2}}$, that is $|y|=\ord{\sqrt\eps}$, then
\begin{itemize}
	\item  the errors in doing this are typically $\ord{G(\Psi)}
	=\ord{\exp(-\beta/\eps)}$, namely exponentially small;
	
	\item  and for large $Y<\Psi$, at worst $\V{n+1}=\ord{\sqrt\eps\V{n}}$
	which preserves the ordering of the asymptotic terms.
\end{itemize}
Consequently, the above expansion is asymptotically valid.

\subsection{Independent jittering}

Similarly, in the {\sde}~(\ref{tcee15}), we would like to replace the factor
$\Z2dW_1$ by a $dw$ where
\[
	dw = \Z2dW_1\,,  \quad
	d\Z2 = -\beta\Z2dt+dW_2\,.
\]
Here the two noise processes, $W_1$ and $W_2$, are independent, thus
causing the Fokker-Planck equation of this Stratonovich {\sde} to be the
somewhat simpler
\begin{equation}
	\partial_t u=
	\beta\partial_y\left(yu\right)
	+\frac{\eps^2}{2}\left\{y^2\partial_{xx}u
	   +\partial_{yy}u \right\}\,,
	\label{fpbb}
\end{equation}
where here $x=w$ and $y=\Z2$.
As before, we seek a model for the
long-term evolution of $x(t)$ by substituting~(\ref{fpcm}--\ref{fpevol})
into this equation  to find
\begin{eqnarray}
	U & \sim & G(Y)p
	\nonumber \\
	 &  & +\eps^4\left(\frac{1}{4\beta}Y^2
	 -\frac{1}{8\beta^2}\right)G(Y)p_{xx}
	\label{fpcm4} \\
	 &  & +\eps^8\left(\frac{1}{32\beta^2}Y^4 +\frac{1}{32\beta^3}Y^2
	 -\frac{5}{128\beta^4}\right)G(Y)p_{xxxx} \,,
	 \nonumber \\
	 \mbox{such that}\quad
	 p_t&\sim& \frac{\eps^4}{4\beta}p_{xx} +\frac{\eps^8}{16\beta^3}p_{xxxx}
	 \,.
	\label{fpev4}
\end{eqnarray}
This last evolution equation, when truncated to neglect terms
$o\left(\eps^4\right)$, is the Fokker-Planck equation of a {\sde}
such as
\begin{equation}
	dx\sim\frac{\eps^2}{2\sqrt{\beta}}dW_4\,,
	\label{fpsde4}
\end{equation}
where the noise process $dW_4$ may be considered essentially independent of
the other noises $dW_1$, $dW_2$ and $dW_3$.  Again numerical experiments
confirm this as seen by the lack of correlations in Figure~\ref{fzdw2}.
\begin{figure}
\centering\includegraphics{zdwa2.ps}
\caption{scatter plots on the
time-scale $\Delta t=2/\beta$: left-hand plot of $\Delta x$ versus $\Delta
W_1$; right-hand plot for $\Delta x$ versus $\Delta W_2$.  As before, the
data is generated by numerical integration over $0\leq t\leq 10,000$ with a
time-step of $0.01$.}
\protect\label{fzdw2}
\end{figure}
Here, this is so even on the relatively short time-scale of $\Delta
t=2/\beta$.  There is no drift term in this {\sde}; the analysis suggests
that over long time-scales we may treat the nonlinear noise $\Z2dW_1$ by a
white noise $\frac{1}{2}dW_4/\sqrt{\beta}$.

\subsection{Simplest, accurate, long-term model}

The preceding analysis suggests that we may write the low-dimensional model
\begin{eqnarray}
ds&\sim&\left[\left(\alpha +\frac{ab}{\beta^2}F_x^2\right)s
-\frac{ab}{\beta}s^3+\frac{2\alpha
ab}{\beta^2}s^3-\frac{2a^2b^2}{\beta^3}s^5\right]dt
\nonumber\\&&
+\left(1+\frac{2ab}{\beta^2}s^2\right)F_x dW_1
-\left(\frac{a}{\beta}s+\frac{4a^2b}{\beta^3}s^3\right)F_y dW_2
\label{sdesimp}\\&&\nonumber
-\frac{a}{2\beta^{3/2}}F_xF_y dW_4
+\frac{ab}{\beta^{5/2}}sF_x^2 dW_3
\,.
\end{eqnarray}
This is the ``simplest'' model because:
\begin{itemize}
\item firstly, it is low-dimensional (here just one-dimensional instead of
the original two-dimensions);
	
\item secondly, it has no fast time-scale processes in it at all---other
than those
we call purely white noises whose long-term behaviour is known---and so may
be numerically integrated using standard methods \cite{Kloeden92} with
large time steps, ones appropriate to the phenomena of interest rather than ones
dictated by the rapid time-scale of negligible modes.
\end{itemize}
Note that this model is a ``weak'' model unlike those of the previous
sections. Previously, long-term fidelity of the model when compared with the
original trajectories is assured by the algebra of the theory.  However,  here
we have resorted, through the Fokker-Planck analysis, to using only broad
statistics of $\Z3 dW_1$ and $\Z2 dW_1$ in the resultant model.  The
short-term details of $\Z3 dW_1$ and $\Z2 dW_1$, which are apparently
needed to maintain fidelity, have been discarded.  However, the processes
$W_1$ and $W_2$ are invoked as noise precisely because we do not know their
detailed dynamics, and so there is no loss in subsequently admitting
that we do not know details of $\Z3 dW_1$ and $\Z2 dW_1$.

Throughout this analysis we have assumed that the original system was
perturbed by white noise.  However, the generalisation to ``coloured''
noise, Ornstein-Uhlenbeck processes, is straightforward, one just adjoins a
dynamical equation describing the evolution of the coloured noise in terms
of a new white noise.  Consequently, we have shown how dynamical systems
perturbed by even coloured noise may be systematically modelled by a
low-dimensional system involving purely white noises.

On a philosophical level, we see in the process of reduction
from~(\ref{eqx}--\ref{eqy}) to (\ref{sdesimp}) how unknown effects,
represented by $W_1$ and $W_2$, effectively give rise to multiple noises in
a mathematical model such as (\ref{sdesimp}).  Moreover, through the
nonlinearities, the originally additive unknown processes naturally generate
multiplicative noises.

\section{Conclusion}

This report has concentrated on a specific toy {\sde}.  This is to
demonstrate how crucial ideas can be developed and threaded together to
form a low-dimensional modelling paradigm for {\sde}s in general.  By
avoiding complicated generalities we have been able to extend the centre
manifold analysis to this interesting class of problems.  Currently we are
using the ideas described herein to investigate physical problems of
significant interest.  For example, we aim to investigate how stochastic
effects, such as those due to turbulence, affect the dispersion of tracer
in channels and pipes.

The analysis performed here has used the Stratonovich calculus for {\sde}s.
As demonstrated by Sch\"oner \& Haken \cite{Schoner87} the Ito analysis
should produce completely equivalent results.

\paragraph{Acknowledgements} We thank Steve Cox for his encouragement and
help in the development of these ideas.  This research is partially
supported by a grant from the Australian Research Council.


\end{document}